\documentclass[lnbip]{svmultln}

\usepackage{todonotes}

\begin{document}

\title{Quality Assurance for AI-based Systems: Overview and Challenges}
 
\author{Michael Felderer\inst{1} \and Rudolf Ramler\inst{2}}

\institute{University of Innsbruck, Austria\\
\email{michael.felderer@uibk.ac.at}	\and
Software Competence Center Hagenberg GmbH (SCCH), Austria \\ 
\email{rudolf.ramler@scch.at}}

\authorrunning{Felderer and Ramler} 
\titlerunning{Quality Assurance for AI}

\maketitle

\begin{abstract}
The number and importance of AI-based systems in all domains is growing. With the pervasive use and the dependence on AI-based systems, the quality of these systems becomes essential for their practical usage. However, quality assurance for AI-based systems is an emerging area that has not been well explored and requires collaboration between the SE and AI research communities. This paper discusses terminology and challenges on quality assurance for AI-based systems to set a baseline for that purpose. Therefore, we define basic concepts and characterize AI-based systems along the three dimensions of artifact type, process, and quality characteristics. Furthermore, we elaborate on the key challenges of (1) understandability and interpretability of AI models, (2) lack of specifications and defined requirements, (3) need for validation data and test input generation, (4) defining expected outcomes as test oracles, (5) accuracy and correctness measures, (6) non-functional properties of AI-based systems, (7) self-adaptive and self-learning characteristics, and (8) dynamic and frequently changing environments.
\end{abstract}


\section{Introduction} \label{sec:intro}

Recent advances in Artificial Intelligence (AI), especially in machine learning (ML) and deep learning (DL), and their integration into software-based systems of all domains raise new challenges to engineering modern AI-based systems. These systems are data-intensive, continuously evolving, self-adapting, and their behavior has a degree of (commonly accepted) uncertainty due to inherent non-determinism. These characteristics require adapted and new constructive and analytical quality assurance (QA) approaches from the field of software engineering (SE) in order to guarantee the quality during development and operation in live environments. However, as pointed out by Borg~\cite{borg2021aiq}, already the concept of "quality" in AI-based systems is not well-defined. Furthermore, as pointed out by Lenarduzzi et al.~\cite{lenarduzzi2021sqai}, terminology differs in AI and software engineering. 

The knowledge and background of different communities are brought together for developing AI-based systems. While this leads to new and innovative approaches, exciting breakthroughs, as well as a significant advancement in what can be achieved with modern AI-based systems, it also fuels the babel of terms, concepts, perceptions, and underlying assumptions and principles. For instance, the term "regression" in ML refers to regression models or regression analysis, whereas in SE it refers to regression testing. Speaking about "testing", this term is defined as the activity of executing the system to reveal defects in SE but refers to the evaluation of performance characteristics (e.g., accuracy) of a trained model with a holdout validation dataset in ML. The consequences are increasing confusion and potentially conflicting solutions for how to approach quality assurance for AI-based systems and how to tackle the associated challenges. While this paper starts from a software engineering point of view, its goal is to incorporate and discuss also many other perspectives, which eventually aggregate into a multi-dimensional big picture of quality assurance for AI-based systems.   

In this paper, we first discuss the terminology on quality assurance for AI in Section~\ref{sec:terminology}. Then, we discuss challenges on QA for AI in Section~\ref{sec:challenges}. Finally, in Section~\ref{sec:conclusion} we conclude the paper.





\section{Background and Terminology}
\label{sec:terminology}

\emph{AI-based system} (also called \emph{AI-enabled system}) refers to a software-based system that comprises AI components besides traditional software components. However, there are different definitions of what AI means, which vary in their scope and level of detail. AI is (human) intelligence demonstrated by machines, which implies the automation of tasks that normally would require human intelligence. For our context, i.e., quality assurance, we pragmatically include those AI techniques in our working definition of AI that require new or significantly adapted quality assurance techniques. This comprises supervised ML and DL, which require to transfer control from source code (where traditional QA can be applied) to data. Borg~\cite{borg2021aiq} explicitly introduces the term \emph{MLware} for the subset of AI that, fueled by data, realizes functionality through machine learning.

\emph{Software quality} is defined as the capability of a software product to satisfy stated and implied needs when used under specified conditions~\cite{ISO2005SoftwareQualityGuide}. \emph{Software quality assurance} is then the systematic examination of the extent to which a software product is capable of satisfying stated and implied needs~\cite{ISO2005SoftwareQualityGuide}.

\emph{AI components}, especially based on supervised ML or DL, differ fundamentally from traditional components because they are data-driven in nature, i.e., their behavior is non-deterministic, statistics-orientated and evolves over time in response to the frequent provision of new data~\cite{amershi2019software}. An AI component embedded in a system comprises the data, the ML model, and the framework. Data are collected and pre-processed for use. Learning program is the code for running to train the model. Framework (e.g., Weka, scikit-learn, and TensorFlow) offers algorithms and other libraries for developers to choose from when writing the learning program.

\begin{figure}[h]
    \centering
    \includegraphics[width=\textwidth]{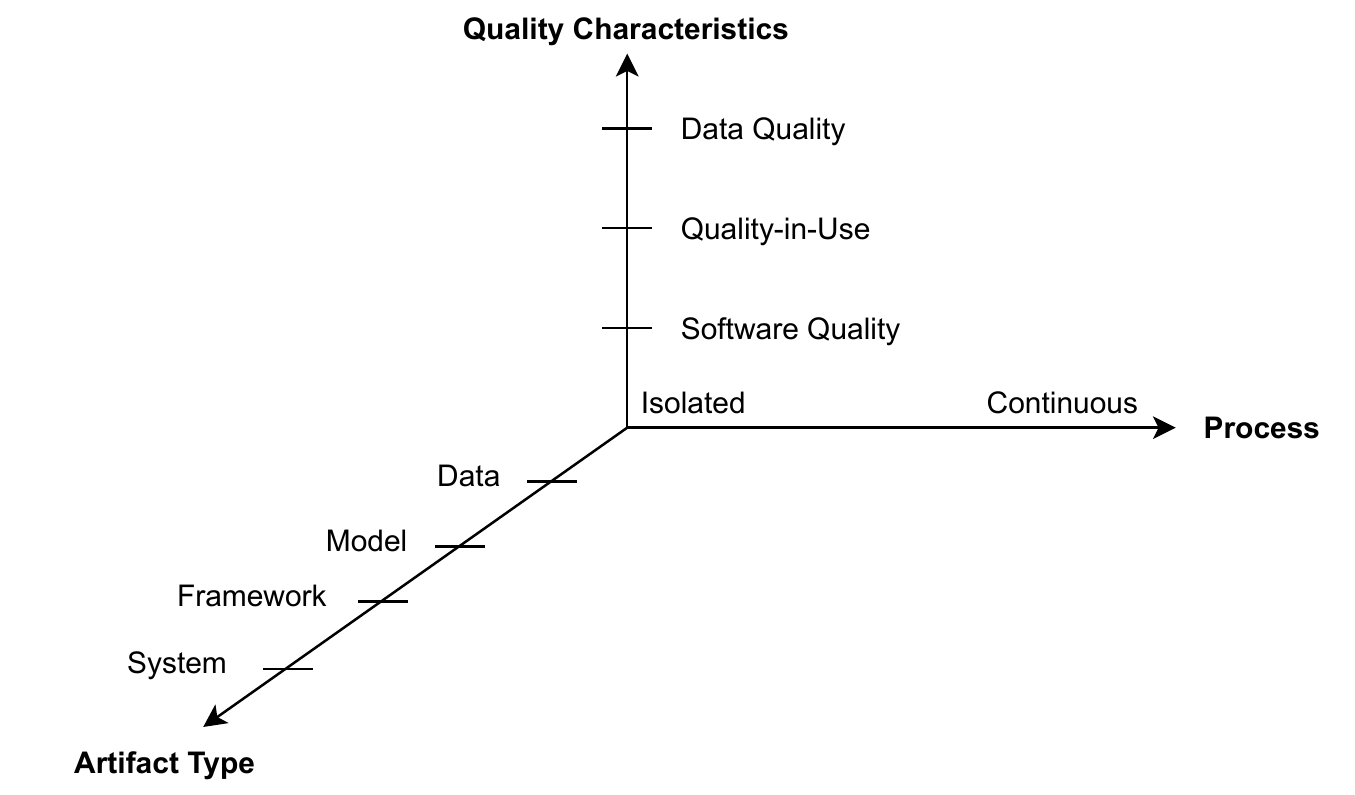}
    \caption{Dimensions of AI-based Systems and Quality Assurance}
    \label{fig:qaai}
\end{figure}

To characterize AI-based systems for the purpose of quality assurance, it is meaningful to consider several dimensions. Such dimensions are the \emph{artifact type} dimension, the \emph{process} dimension and the \emph{quality characteristics} dimension. The dimensions and their values are shown in Fig.~\ref{fig:qaai}.

On the artifact type dimension, we can consider, based on the characterization of AI components in the previous paragraph, the system, framework, model and data perspective. On the process dimension, we can distinguish whether AI components and systems are developed in isolation or continuously by iteratively taking feedback from the deployed components into account based on DevOps principles. For all artifact and process settings, quality characteristics are relevant.

For instance, additional \emph{quality properties} of AI components and AI-based systems have to be taken into account. Zhang et al.~\cite{zhang2020machine} consider the following quality properties:
\begin{itemize}
    \item \emph{Correctness} refers to the probability that an AI component gets things right.
    \item \emph{Model relevance} measures how well an AI component fits the data.
    \item \emph{Robustness} refers to the resilience of an AI component towards perturbations.
    \item \emph{Security} measures the resilience against potential harm, danger or loss made via manipulating or illegally accessing AI components. 
    \item \emph{Data privacy} refers to the ability of an AI component to preserve private data information.
    \item \emph{Efficiency} measures the construction or prediction speed of an AI component.
    \item \emph{Fairness} ensures that decisions made by AI components are in the right way and for the right reason to avoid problems in human rights, discrimination, law, and other ethical issues.
    \item \emph{Interpretability} refers to the degree to which an observer can understand the cause of a decision made by an AI component.
\end{itemize}

Felderer et al.~\cite{felderer2019testing} highlight the additional importance of \emph{data quality} for the quality of AI components. According to ISO/IEC 25012~\cite{ISO2008DataQuality} data quality characteristics in the context of software development can be classified into inherent and system-dependent data characteristics. \emph{Inherent data quality} refers to data itself, in particular to data domain values and possible restrictions, relationships of data values and meta-data. \emph{System-dependent data quality} refers to the degree to which data quality is reached and preserved within a system when data is used under specified conditions. For the framework, which ultimately is software, the classical software quality characteristics based on ISO/IEC 25010~\cite{ISO2011SoftwareQuality}, i.e., effectiveness, efficiency, satisfaction, freedom from risk and context coverage for quality in use as well as functional suitability, performance efficiency, compatibility, usability, reliability, security, maintainability, and portability for system/software product quality can be applied.

\emph{Testing of AI components} or \emph{AI-based systems} refers to any activity aimed at detecting differences between existing and required behaviors of AI components or AI-based systems. The testing properties (such as correctness, robustness, or efficiency) stem from the quality characteristics defined before. Testing can target the data, the ML model, the framework, or the entire system.

Depending on whether testing activities for AI components are performed before or after ML model deployment one can distinguish offline and online testing. \emph{Offline testing} tests the AI component with historical data, but not in an application environment~\cite{zhang2020machine}. Cross-validation using a validation dataset is a typical offline testing approach to make sure that the AI component meets the required conditions. \emph{Online testing} tests deployed AI components in a real or virtual application environment. Online testing complements offline testing, because the latter relies on historical data not fully representing future data, is not able to test some problematic circumstances occurring in real environments like data loss, and has no access to direct user feedback. A common online testing technique is \emph{A/B testing}, which is a splitting testing technique to compare two or more versions of a deployed component. A/B tests are often performed as experiments and the activity is called continuous experimentation in software engineering~\cite{ros2018continuous,auer2018current}.


\section{Challenges} \label{sec:challenges}

A wide range of challenges exists, which stem from the novelty of the topic. Currently, there is a lack of (standardized) approaches for quality assurance of AI-based systems. Many attempts are in progress to fill the void. Yet the understanding of the problem is still very incomplete. It prolongs to fundamental questions like what are relevant quality characteristics (see previous section) and what is a bug. An example for a "new type of bug" unseen in conventional software is the phenomenon of adversarial examples \cite{yuan2019adversarial}, where small variations in the input (e.g., noise in image data or recorded speech that is not or barely noticeable for the human user) has a dramatic impact on the output as it results in a severe misclassification. 

In addition to outlining important concepts and terms in the previous section, this section elaborates on the following key challenges encountered in the development of approaches for quality assurance and testing of AI-based systems.
\begin{itemize}
    \item Understandability and interpretability of AI models
    \item Lack of specifications and defined requirements
    \item Need for validation data and test input generation
    \item Defining expected outcomes as test oracles
    \item Accuracy and correctness measures
    \item Non-functional properties of AI-based systems
    \item Self-adaptive and self-learning characteristics
    \item Dynamic and frequently changing environments
\end{itemize} 

\subsubsection{Understandability and interpretability:} Data scientists are struggling with the problem that ML and in particular DL are producing models that are opaque, non-intuitive, and difficult for people to understand. The produced models turned out to be uninterpretable "black boxes" \cite{goebel2018explainable}. This challenge propagates to testing and quality assurance activities and it affects debugging models when they have confirmed defects. Black-box testing is a common approach in software quality assurance. So why does the lack of understandability and interpretability also have an impact on testing? 
The challenge for quality assurance results from the \emph{lack of specifications and defined requirements} that developers and testers are used to have for conventional software systems and which provide the knowledge necessary to understand, build and test the system~\cite{bosch2018takes}.

\subsubsection{Lack of specifications and defined requirements:}
Data-based/learning-based approaches do not rely on specifications and predefined requirements. They automatically generate models from existing data. The data used for learning consists of a wide range of input and labeled output. Model generation is an exploratory approach. Learning algorithms are applied to seek relevant "rules" how to connect the input to the expected output. Whether such rules can be found and how adequate they are to accurately model the connection is usually unclear at the beginning of the learning process. 

Conventional software development works in the exact opposite way compared to data-based/learning-based approaches~\cite{fischer2020applying}. Specifications are defining the required behavior of the system, i.e., the "rules". They are available before the system is implemented. People have learned about relevant rules, for example, by experience (e.g., domain experts) or because they have acquired the knowledge from specifications (e.g., developers). 
The goal in testing conventionally developed systems is to come up with inputs and labeled outputs to verify and validate the implemented rules. Testing explores representative scenarios as well as boundaries and corner cases. This goal is also important for testing AI-based systems. However, testing techniques for conventional systems are supposed to rely on specifications to derive inputs or to determine the expected outcome for an input, which leads to further challenges such as the challenge of \emph{test input generation} and \emph{defining test oracles} when testing AI-based systems.

\subsubsection{Test input generation:} 
In testing, it is usually the case that systems have a huge input space to be explored. Hence, at the core of any testing approach is the problem that completely exercising even a moderately complex system is impossible due to the combinatorial explosion of the number of possible inputs. Testing AI-based systems is no difference~\cite{marijan2019challenges}. 

Software testing techniques commonly deal with the challenge of huge input spaces by adopting sampling strategies for selecting inputs when designing test cases. A number of testing techniques have been developed that are classified~\cite{7346375} as specification-based (black-box), structure-based (white-box), or experience-based. 
Similar techniques suitable for AI-based system testing are yet to emerge. First techniques have been proposed that exploit structure information of deep neural networks to derive coverage measures such as various forms of neuron coverage (see, e.g.,~\cite{10.1145/3293882.3330579}). Inputs (test data) is generated with the goal to maximize coverage. Various approaches are currently explored, from random generation (fuzzing)~\cite{10.1145/3293882.3330579} to GAN-based metamorphic approaches~\cite{zhang2018deeproad}. However, due to the lack of interpretability and understandability (resulting from a lack of specifications and requirements), identifying and selecting representative inputs to construct meaningful test cases is still an open challenge~\cite{braiek2020testing}. 

\subsubsection{Defining test oracles:} 
The goal of testing is to reveal faults in terms of incorrect responses or behavior of the system in reaction to a specific input. In order to determine whether the observed output (responses and behavior) is correct or incorrect, it has to be compared to some expected output. The source providing information about what is a correct output is called test oracle~\cite{barr2015oracle}. In manually constructing test cases, a human tester defines the input and the corresponding expected output. In a production setting, however, the input is dynamically created throughout the actual use of the system in a particular context or environment. It typically includes values and value combinations that have never been used before and which were even not anticipated to be used at all. Hence, the "oracle problem" of determining the correct output for an input, a core challenge in testing, dramatically increases when testing in performed in production environments under diverse settings.

\subsubsection{Accuracy and correctness:} 
Closely related is the \emph{accuracy problem}. Software is expected to be deterministic and correct. Any deviation from the expected behavior and any difference in the output is considered a defect that is supposed to be fixed. It is well known, that real-world software is not defect-free and there is no perfect system. However, the underlying principles and the level of correctness currently achieved in software engineering is different from what AI-based systems exhibit. AI-based systems are accepted to be inherently "defective", as they usually operate in a defined accuracy range. Yet a system with 99\% accuracy will "fail" in about one out of hundred runs. 
Applying conventional testing and quality assurance principles and approaches is incompatible with the underlying assumption that the system is considered correct although it exhibits a high number of contradicting ("failing") cases. The corresponding testing techniques and quality metrics developed for deterministic systems first need to be adapted before they can be used to assess systems with probabilistic behavior.


\subsubsection{Non-functional properties:}
Testing for \emph{non-functional aspects} is always challenging and requires suitable approaches to specify expected properties. This also holds for testing of AI-based systems, where testing non-functional aspects has rarely been explored~\cite{zhang2020machine}. 
Especially robustness and efficiency are well suited for testing in production. Testing robustness of AI components is challenging because input data has more diverse forms, for instance image or audio data. Especially adversarial robustness, where perturbations are designed to be hard to detect are also hard to define in terms of corresponding test oracles. Metamorphic relations~\cite{xie2011testing,dwarakanath2018identifying} are therefore frequently exploited as alternative ways to construct test oracles.
Testing efficiency for AI components has to deal not only with prediction speed, but also with construction speed, which poses challenges to measuring and analyzing performance, especially in a real-time context when decisions have to be made instantaneous (e.g., in autonomous driving). 

\subsubsection{Self-adaptive and self-learning systems:}
Regression testing is a major task in any modern software development project. The agile paradigm and the DevOps movement have led to short development cycles with frequent releases as well as the widespread use of techniques such as Continuous Integration, Deployment, and Delivery~\cite{humble2010continuous}. The answer to the question how quality assurance can keep up with the continuously growing development speed is automated testing. Test automation, however, is a major cost-driver. First, due to the effort for initially setting up the test environment and implementing the automated tests, and second, even more so due to the effort for maintaining the automated tests when the system under test has been changed~\cite{garousi2016developing}. In contrast to conventional software that is evolved over a series of development iterations, many AI-based systems are designed to evolve dynamically at run-time by self-adapting to changing environments and continuously learning from the input they process~\cite{khritankov2021feedbackloops}. Testing dynamic self-adaptive systems raises many open issues about how to cope with the complexity and dynamics that result from the flexibility of self-adaptive and self-learning systems~\cite{eberhardinger2014towards}.

\subsubsection{Dynamic environments:}
AI components often operate in dynamic and frequently changing environments. Examples are typically data intensive applications that have to integrate data from various sources (including sensors, web data, etc.), which all have different characteristics regarding their data quality~\cite{foidl2019technical,foidl2019risk}. Data can also stem from simulators or AI components may have to control simulations. Due to the complexity and non-determinism of the environment, testability (i.e., controllability and observability) is highly challenging. Furthermore, due to information heterogeneity also privacy and security aspects are essential. 
To address these issues, run-time monitoring and online testing have been suggested. Online testing, the assessment of the system's behavior is performed live, in production and in a real application environment~\cite{zhang2020machine}. 

Real application environments provide the advantage of real user integration and real user experience. In modern cloud-based environments user information can easily be collected and used to evaluate and continuously improve the system (e.g., in web-based recommender systems). However, this requires a significant number of users with a clear user profile. In addition, applying testing in production for business-critical users poses business risks. In addition, one has to carefully select metrics to guarantee their validity and reliability. 
The term "testing in production" can even be considered as an oxymoron, especially if systems are safety-critical and can harm the health of impacted stakeholders (e.g., for autonomous systems controlling vehicles). In that context, clear constraints have to be defined and guarantees under which conditions testing in production can be performed at all because safety-criticality requires clear strategies to remove defects before deployment or to handle them properly in production. However, besides safety also privacy and ethical issues may restrict the applicability of testing in production and therefore require specific constraints and monitors.

\section{Summary and Conclusions} \label{sec:conclusion}

In this paper, we discussed terminology and challenges on quality assurance for AI-based systems. To characterize AI-based systems for the purpose of quality assurance, we defined the three dimensions artifact type (i.e., data, model, framework, and system), process (from isoloated to continuous), and quality characteristics (with respect to software quality, quality-in-use, and data quality). Furthermore, we elaborated on the key challenges of (1) understandability and interpretability of AI models, (2) lack of specifications and defined requirements, (3) need for validation data and test input generation, (4) defining expected outcomes as test oracles, (5) accuracy and correctness measures, (6) non-functional properties of AI-based systems, (7) self-adaptive and self-learning characteristics, and (8) dynamic and frequently changing environments.

In order to properly address the challenges raised in this paper and to enable high quality AI-based systems, first and foremost, exchange of knowledge and ideas between the SE and the AI community is needed. One channel of exchange is education or training through dedicated courses~\cite{kastner2020teaching} or media~\cite{hulten2019building}. Another one are dedicated venues for exchange and discussion of challenges on quality assurance for AI-based systems like the IEEE International Conference On Artificial Intelligence Testing or the workshop Quality Assurance for AI collocated with the Software Quality Days.

\section*{Acknowledgements} \label{sec:acknowledgements}
The research reported in this paper has been partly funded by the Federal Ministry for Climate Action, Environment, Energy, Mobility, Innovation and Technology (BMK), the Federal Ministry for Digital and Economic Affairs (BMDW), and the Province of Upper Austria in the frame of the COMET - Competence Centers for Excellent Technologies Programme managed by Austrian Research Promotion Agency FFG.

\bibliographystyle{splncs}
\bibliography{references}

\end{document}